\documentclass[twoside,twocolumn]{article}

\usepackage{blindtext} % Package to generate dummy text throughout this template 

\usepackage[sc]{mathpazo} % Use the Palatino font
\usepackage[T1]{fontenc} % Use 8-bit encoding that has 256 glyphs
\usepackage{microtype} % Slightly tweak font spacing for aesthetics

\usepackage[english]{babel} % Language hyphenation and typographical rules

\usepackage[hmarginratio=1:1,top=32mm,columnsep=20pt]{geometry} % Document margins
\usepackage[hang, small,labelfont=bf,up,textfont=it,up]{caption} % Custom captions under/above floats in tables or figures
\usepackage{booktabs} % Horizontal rules in tables

\usepackage{lettrine} % The lettrine is the first enlarged letter at the beginning of the text

\usepackage{enumitem} % Customized lists
\setlist[itemize]{noitemsep} % Make itemize lists more compact

\usepackage{abstract} % Allows abstract customization
\usepackage{titlesec} % Allows customization of titles
\renewcommand\thesection{\Roman{section}} % Roman numerals for the sections
\renewcommand\thesubsection{\roman{subsection}} % roman numerals for subsections
\titleformat{\section}[block]{\large\scshape\centering}{\thesection.}{1em}{} % Change the look of the section titles
\titleformat{\subsection}[block]{\large}{\thesubsection.}{1em}{} % Change the look of the section titles

\usepackage{fancyhdr} % Headers and footers
\usepackage{titling} % Customizing the title section

\usepackage{hyperref} % For hyperlinks in the PDF
\usepackage{natbib}
\usepackage{amsmath,amssymb,amsthm,bm}
\usepackage{color, graphics, graphicx}
\usepackage{cases}
\pretitle{\begin{center}\Huge\bfseries} % Article title formatting
\posttitle{\end{center}} % Article title closing formatting
\title{On the quasi-static effective behaviour of  poroelastic media containing elastic inclusions} % Article title
\author{%
\textsc{Pascale Royer, Pierre Recho, Claude Verdier} 
}
\date{} % Leave empty to omit a date
\renewcommand{\maketitlehookd}{%
\begin{abstract}
The aim of the present study is to derive the effective quasi-static behaviour of a  composite medium, made of a poroelastic matrix containing elastic impervious inclusions. For this purpose, the asymptotic homogenisation method is used. On the local scale, the governing equations include Biot's model of poroelasticity in the porous matrix and Navier equations in the inclusions, with elastic properties of the same order of magnitude.  Biot's diphasic model of poroelasticity is obtained on the macroscopic scale, but with effective parameters that are strongly impacted by the distribution of inclusions, even at low volume fraction. The impact on fluid flow is strictly geometrical, showing that the inclusions do not play the role of a porous network.
\end{abstract}
}

\begin{document}

% Print the title
\maketitle
\section{Introduction}
Composites made of a porous matrix reinforced with solid impervious inclusions  occur in several engineering disciplines, involving natural media such as geomaterials \cite{Rice78}, biological tissues \cite{Rau18}, \cite{Lob03}, or tumors \cite{Xue17}, as well as man-made structures as, cement-based \cite{Lem02} or biomimetic materials \cite{Raj10}. A comprehensive  understanding of the overall behaviour of these composites, on the basis of their microstructure, can enhance the knowledge concerning  physical scenarios,  with respect to key physi\-cal properties. In this way, a generalisation of Esh\-elby's formula is proposed in \cite{Berr97}, to give the response of a single ellipsoidal elastic inclusion, in a poroelastic whole space, to a uniform strain imposed at infinity.
Some other works focus on fluid flow or solute transport, such as the mathematical model developed in \cite{Fed08}, of the effect of fibre arrangement on the permeability of a porous fibre-reinforced composite, or the macroscopic models  obtained in \cite{Bal03} by asymptotic homogenisation, for passive solute transport in a rigid medium made of a porous matrix with impervious inclusions. 

The present work is aimed at deriving the effective quasi-static mechanical behaviour of a sa\-turated poroelastic medium containing elastic impervious inclusions and is focused  on constituents with elastic properties of the same order of ma\-gnitude. For this purpose, the method of asymptotic homogenisation is used and Biot's model of poroelasticity  is obtained on the macroscopic scale, but with effective parameters that are strongly affected by the distribution of inclusions. The paper is organised as follows. Section \ref{hom} presents a brief description of the homogenisation methodo\-logy. Then, homogenisation of a poroelastic medium with elastic inclusions is detailed in Section \ref{main_sec}, and the derived macroscopic description is commented in Section \ref{macro_desc}. Finally, Section \ref{conclu} presents a summary of the main theoretical results contained in this work and highlights conclusive remarks.
\section{Homogenisation method}
\label{hom}
\subsection{Medium under consideration}
\begin{figure*}[ht!!]
    \begin{center}
    \includegraphics[width=7cm]{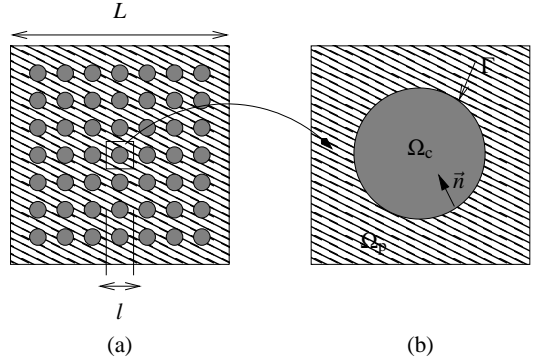}
    \caption{\label{milieu}{\it\small Porous medium :(a) Macroscopic sample; (b) Periodic unit cell.}}
    \end{center}
\end{figure*}

We consider a periodic medium, of characteristic size $L$, and  made of a fluid saturated elastic porous matrix  which contains isolated elastic impervious inclusions. We further denote the spacing between two inclusions by $l$, and and we formulate the condition of separation of scales by $\varepsilon = {l}/{L} \ll 1$.
Within the periodic cell $\Omega$, we denote by $\Omega_\mathrm{p}$ the fluid-saturated porous matrix domain, by $\Omega_\mathrm{c}$ the volume occupied by the inclusion, and by $\Gamma$ their common interface, as depicted in Fig.\ref{milieu}.
Using the two characteristic lengths, $l$ and $L$, and the physical space variable, $\vec{X}$, 
we define two dimensionless space variables:
$ \vec{y} = {\vec{X}/}{l}$, $\vec{x} = {\vec{X}}/{L}$, and $\vec y$ and $\vec x$ describe variations on the microscopic and the macroscopic scales, respectively.
Invoking the differentiation rule of multiple variables, the gra\-dient operator with respect to $\vec{X}$  is written as
\begin{equation}
    \vec\nabla_X = \dfrac{1}{l}\vec\nabla_y + \dfrac{1}{L}\vec\nabla_x.
    \label{grad_dim}
\end{equation}
We further introduce the following cell averages
$$
\begin{array}{l}
<. >^{\Omega}=<.>^{\Omega_\mathrm p}+<.>^{\Omega_\mathrm c},\vspace{0,2cm}\\
<.>^{\Omega_\mathrm \alpha}=
\displaystyle\dfrac{1}{\mid\Omega\mid}\int_{\Omega_\alpha}\ .\ d\Omega
\hspace{0,5cm}(\alpha=\mathrm p, \mathrm c).
\end{array}
$$
\subsection{Homogenisation procedure}
The methodology firstly consists in writing, in dimensionless form,
the governing equations which describe the problem on the periodic unit cell.
This dimensionless writing of the equations requires the choice of a characteristic length 
for the dimensionless writing of space derivatives. We arbitrarily choose $L$ as the reference characteristic length. 
The dimensionless gradient operator is thus $L \vec\nabla_X $, which by Eq.(\ref{grad_dim}) is given by
\begin{equation}
    \vec\nabla = L \vec\nabla_X = \varepsilon^{-1} \vec\nabla_y + \vec\nabla_x.
\label{dimensionless_grad}
\end{equation}
The homogenisation method being used is based upon the fundamental assumption
that the unknown fields can be written in the form of asymptotic expansions in powers of $\varepsilon$
\begin{equation}
	\psi = \psi^{0}\left(\vec y, \vec x\right) + \varepsilon \psi^{1}\left(\vec y, \vec x\right) + \varepsilon^2 \psi^{2} \left(\vec y, \vec x\right) +...,
	\label{dev_asymp}
\end{equation}
in which functions $\psi^{i}$ are $\Omega$-periodic in variable $\vec y$.
The method consists in incorporating the asymptotic expansions in the dimensionless local description,
while taking into account the expression of the dimensionless gradient operator 
Eq.(\ref{dimensionless_grad}).
 This leads to approximate governing equations and boundary conditions at the successive orders, 
which together with the condition of periodicity define well posed boundary value problems within the periodic unit cell, from which functions $\psi^i$ can be determined.
 Existence of solutions requires that volume ave\-raged equations be satisfied. 
The latter ones actually describe the macroscopic behaviour at successive orders.
\section{Quasi-static homogenisation in a poro\-elastic medium with elastic inclusions}
\label{main_sec}
\subsection{Dimensionless governing equations\\ on the local scale}
The poroelastic matrix ($\Omega_\mathrm p $)  is made of a li\-near elastic skeleton saturated with a viscous incompressible Newtonian fluid, and its behaviour is  described by quasi-static Biot's model \cite{Biot55}\footnote{This is possible provided that the matrix pore size, $l_\mathrm p$ be greatly smaller than the spacing between two inclusions, $l$. }:
\begin{numcases}{}
 \vec\nabla\cdot\tilde{\sigma}_\mathrm p=\vec 0,\label{eqp1}\vspace{0,3cm}\\
 \tilde{\sigma}_\mathrm p=\tilde {c}_\mathrm p:\tilde e(\vec{u}_\mathrm s)-\tilde{\alpha}_\mathrm p\ p_\mathrm f,\label{eqp2}\vspace{0,2cm}\\
 \vec\nabla\cdot\vec{v}_\mathrm p = - \tilde\alpha_\mathrm p:\ \tilde e (\dfrac{\partial \vec { u}_\mathrm s}{\partial t}) - \beta_\mathrm p \dfrac{\partial p_\mathrm f}{\partial t},\label{eqp3}\\
 \vec {v}_\mathrm p=\phi_\mathrm p (\vec {v}_\mathrm f - \dfrac{\partial \vec {u}_\mathrm s}{\partial t})=- \dfrac{\tilde {K}_\mathrm p}{\mu}\vec\nabla p_\mathrm f.\label{eqp4}
\end{numcases}
The four above equations express the momentum balance, the poroelastic constitutive law, the conservation of fluid mass
and Darcy's law, respectively. The distinct quantities involved in the model are the following: 
$\tilde{\sigma}_\mathrm p $ and $p_\mathrm f$  denote the total stress and the interstitial fluid pressure, respectively;
$\vec {u}_\mathrm s$ is the solid displacement of the porous matrix, while $\vec {v}_\mathrm f$ and $\vec {v}_\mathrm p$ stand for the mean fluid velocity within the volume of the micropores and the mean fluid relative velocity within the porous matrix;
$ \tilde e({\vec{u}}_\mathrm s)=1/2(\vec\nabla\vec u_\mathrm s+\vec\nabla^{\mathrm{T}}\vec u_\mathrm s)$ is the strain tensor, while ${\tilde c}_\mathrm p$, $\tilde{\alpha}_\mathrm p$, $\beta_\mathrm p>0$, and $\tilde {K}_\mathrm p $ represent  the fourth order elastic tensor of the drained porous matrix, the second order symmetric and positive Biot coupling tensor, Biot's bulk modulus and the second order tensor of permeability of the porous matrix, respectively;
$\phi_\mathrm p$ and $\mu$ denote the porosity of the porous matrix, and the fluid visco\-sity, respectively.

The inclusion ($\Omega_\mathrm c $) is linear elastic and satisfies the Navier equations 
\begin{numcases}{}
  \vec\nabla\cdot\tilde{\sigma}_\mathrm c=\vec 0,\label{eqc1}\\
  \tilde{\sigma}_c=\tilde {c}_c:\tilde e(\vec{u}_\mathrm c),\label{eqc2}
\end{numcases}
where ${\tilde\sigma}_\mathrm c $ and $\vec{u}_\mathrm c$ stand for the solid stress tensor and displacement, respectively, and where $\tilde {c}_\mathrm c$ represents the elastic tensor.

The appropriate conditions over the interface $\Gamma $ between the porous matrix and the inclusion include the continuity of normal stresses and displacements and the normal mean fluid relative velocity within the porous matrix must be set to zero \cite{Mik12}:
\begin{numcases}{}
 \tilde{\sigma}_\mathrm p\cdot\vec n=\tilde{\sigma}_\mathrm c\cdot\vec n
 \hspace{0,5cm}\hbox{over $\Gamma$},\label{eqg1}\\
 \vec {u}_\mathrm s=\vec {u}_\mathrm c
 \hspace{0,5cm}\hbox{over $\Gamma$},\label{eqg2}\\
 \vec {v}_\mathrm p\cdot \vec n =0
\hspace{0,5cm} \hbox{over $\Gamma$}\label{eqg3},
\end{numcases}
where $\vec n$ denotes the unit vector giving the normal to $\Gamma$ exterior to $\Omega_\mathrm p$.
 \subsection{Homogenisation}
 We consider equations Eqs.(\ref{eqp1})-(\ref{eqc2}) and boundary conditions Eqs.(\ref{eqg1})-(\ref{eqg3}), and we look for solutions in the form of  Eq.(\ref{dev_asymp}) for $\tilde\sigma_\mathrm p$, $\tilde\sigma_\mathrm c$, $\vec u_\mathrm s$, $\vec u_\mathrm c$, $p_\mathrm f$ and $\vec v_\mathrm p$. Note that, due to Eq.(\ref{dimensionless_grad}), the expansion of $\vec v_\mathrm p$ starts with a term in $ \varepsilon^{-1}$. Furthermore, the strain tensors read
\begin{equation}
\tilde e(\vec u_\alpha)=\varepsilon^{-1}\tilde e_y(\vec u_\alpha) + \tilde e_x(\vec u_\alpha), \hspace{0,5cm}(\alpha=\mathrm p, \mathrm c),
\label{strain_tens}
\end{equation}
 and consequently the expansions of both stress tensors also start with a $\varepsilon^{-1}$ term.
 Incorporating the asymptotic expansions and the expressions of the dimensionless gradient operator Eq.(\ref{dimensionless_grad}) and of the strain tensors Eq.(\ref{strain_tens})
into Eqs.(\ref{eqp1})-(\ref{eqg3}), then identifying terms of same
power of $\varepsilon$, leads to boundary value problems at the successive orders.
\subsubsection{Boundary value problem for $\tilde \sigma_\mathrm p^{-1} $, $\tilde \sigma_\mathrm c^{-1} $, $\vec u_\mathrm s^0$ and $\vec u_\mathrm c^0$}
Considering the leading order of Eqs.(\ref{eqp1})-(\ref{eqp2}) and Eqs.(\ref{eqc1})-(\ref{eqg2}), we deduce
the following  boundary value problem of unknowns $\vec u_\mathrm s^0$ and $\vec u_\mathrm c^0$:
\begin{numcases}{}
 \dfrac{\partial}{\partial y_j}
 \left[ 
 c_{\mathrm p_{ijlm}}e_{y_{lm}}(\vec u_\mathrm s^0)
 \right]=0\ \hbox{within $\Omega_\mathrm p,$}\\
 \dfrac{\partial}{\partial y_j}
 \left[ 
 c_{\mathrm c_{ijlm}}e_{y_{lm}}(\vec u_\mathrm s^0)
 \right]=0\ \hbox{within $\Omega_\mathrm c ,$}\\
 \strut\left[ c_{\mathrm p_{ijlm}}e_{y_{lm}}(\vec u_\mathrm s^0)\right]n_j=\nonumber\\
 \left[ c_{\mathrm c_{ijlm}}e_{y_{lm}}(\vec u_\mathrm c^0)\right]n_j\ 
 \hbox{over $\Gamma$},\\
 u_{\mathrm s_i}^0=u_{\mathrm c_i}^0\hspace{0,5cm}\hbox{over $\Gamma$},\\
 \tilde \sigma_\mathrm p^{-1}, \tilde \sigma_\mathrm c^{-1}, \vec u_\mathrm s^0, \vec u_\mathrm c^0:\hbox{periodic in $\vec y$},
\end{numcases}
from which it is clear that the displacements $\vec u_\mathrm s^0$ and $\vec u_\mathrm c^0$ are constant over the period
\begin{equation}
 \vec u_\mathrm s^0=\vec u_\mathrm c^0=\vec u^0 (x).
\end{equation}
Since by Eqs.(\ref{eqp2}) and (\ref{eqc2}) at ${\mathcal O}(\varepsilon^{-1})$
\begin{equation}
 \sigma^{-1}_{\alpha_{ij}} = c_{\mathrm p_{ijlm}}e_{y_{lm}}(\vec u^0)
 \hspace{0,5cm}(\alpha=\mathrm p, \mathrm s),
\end{equation}
we consequently get
\begin{equation}
 \tilde\sigma_\mathrm p^{-1}=\tilde\sigma_\mathrm c^{-1}=\tilde 0.
\end{equation}
\subsubsection{Boundary value problem for $\vec v_p^{-1}$ and $p_\mathrm f^0$}
At the lowest order, Eqs.(\ref{eqp3})-(\ref{eqp4}) lead to
\begin{numcases}{}
\dfrac{\partial}{\partial y_i}
\left[\dfrac{K_{\mathrm p_{ij}}}{\mu}\dfrac{\partial p_\mathrm f^0}{\partial y_j} \right]
=0
  \hspace{0,5cm}\hbox{within $\Omega_\mathrm p $},\\
 \left[\dfrac{K_{\mathrm p_{ij}}}{\mu}\dfrac{\partial p_\mathrm f^0}{\partial y_j} \right]n_j=0
 \hspace{0,5cm}\hbox{over $\Gamma$},\\
  p_\mathrm f^0:\hbox{periodic in $\vec y $}.
\end{numcases}
Consequently, we get:
\begin{numcases}{}
  p_\mathrm f^0 =p_\mathrm f^0 (\vec x),\\
  \vec v_\mathrm p^{-1}=\vec 0.
\end{numcases}
\subsubsection{Boundary value problem for $\tilde \sigma_\mathrm p^{0} $, $\tilde \sigma_\mathrm c^{0} $, $\vec u_\mathrm s^1$\\ and $\vec u_\mathrm c^1$}
We now consider the second order of 
Eqs.(\ref{eqp1})-(\ref{eqp2}) and Eqs.(\ref{eqc1})-(\ref{eqg2}), from which we deduce the following system of unknowns 
$\vec u_\mathrm s^1$ and $\vec u_\mathrm c^1$:
\begin{numcases}{}
 \dfrac{\partial}{\partial y_j}\left[c_{\mathrm p_{ijlm}}[e_{y_{lm}}(\vec u_\mathrm s^1)+e_{x_{lm}}(\vec u_\mathrm s^0)]-\alpha_{\mathrm p_{ij}}\ p_f^0  \right]=0\nonumber\\
\hbox{within $\Omega_\mathrm p $},\label{pb1_eq1}\\
\dfrac{\partial}{\partial y_j}\left[ c_{\mathrm c_{ijlm}}[e_{y_{lm}}(\vec u_\mathrm c^1)+e_{x_{lm}}(\vec u_\mathrm c^0)] \right]=0\nonumber\\
\hbox{within $\Omega_\mathrm c $},\\
(c_{\mathrm p_{ijlm}}[e_{y_{lm}}(\vec u_\mathrm s^1)+e_{x_{lm}}(\vec u_\mathrm s^0)]-\alpha_{\mathrm p_{ij}}\ p_f^0)\ n_j=\nonumber\\
(c_{\mathrm c_{ijlm}}(e_{y_{lm}}(\vec u_\mathrm c^1)+e_{x_{lm}}(\vec u_\mathrm c^0)))\ n_j\ 
\hbox{over $\Gamma$},\\
   u_{\mathrm s_i}^1=u_{\mathrm c_i}^1\hspace{0,5cm}\hbox{over $\Gamma$},\label{pb1_eq4}\\
\vec u_\mathrm s^1, \vec u_\mathrm c^1:\hbox{periodic in $\vec y$}.\label{pb1_eq5}
\end{numcases} 
The above set of equations constitutes a well-posed problem for $\vec u_\mathrm s^1$ and $\vec u_\mathrm c^1$, and by virtue of linearity, the solutions read ({\em Cf.} \ref{bvpu1}):
\begin{numcases}{}
  u_{\mathrm s_i}^1=\omega_{\mathrm p_i}^{kh}e_{x_{kh}}(\vec u^0)-\pi_{\mathrm p_i}\ p_\mathrm f^0 + \bar u_{\mathrm s_i}^1 (\vec x),\label{us1}\\
  u_{\mathrm c_i}^1=\omega_{\mathrm c_i}^{kh}e_{x_{kh}}(\vec u^0)+ \bar u_{\mathrm c_i}^1 (\vec x),\label{uc1}
\end{numcases}
where $\bar u_{\mathrm s_i}^1 (\vec x)$ and $\bar u_{\mathrm c_i}^1 (\vec x)$ are arbitrary functions. 
Note that, to render the solution unique, we impose that $\tilde \omega_\mathrm p$, $\vec\pi_\mathrm p$ and $\tilde \omega_\mathrm c$ are with zero average  \cite{Ben78}, \cite{San80}:
$$
<\omega_{\mathrm p_i}^{kh} >^{\Omega_\mathrm p}= 0,\ <\omega_{\mathrm c_i}^{kh} >^{\Omega_\mathrm c}=0,\ <\pi_{\mathrm p_i} >^{\Omega_\mathrm p}=0.
$$
By Eqs.(\ref{eqp2}) and (\ref{eqc2}) at ${\mathcal O}(\varepsilon^{0})$, we obtain
\begin{numcases}{}
 \sigma^{0}_{\mathrm p_{ij}} = 
c_{\mathrm p_{ijlm}}[e_{y_{lm}}(\vec u_\mathrm s^1)+e_{x_{lm}}(\vec u_\mathrm s^0)]\nonumber\\-\alpha_{\mathrm p_{ij}}\ p_f^0,\\
\sigma^{0}_{\mathrm c_{ij}} = c_{\mathrm c_{ijlm}}[e_{y_{lm}}(\vec u_\mathrm c^1)+e_{x_{lm}}(\vec u_\mathrm c^0)],
\end{numcases}
and then employing Eqs.(\ref{us1})-(\ref{uc1}), we deduce
\begin{numcases}{}
 \sigma_{\mathrm p_{ij}}^0=
 (c_{\mathrm p_{ijlm}}e_{y_{lm}}(\vec\omega^{kh})+c_{\mathrm p_{ijkh}})
  e_{x_{kh}}(\vec u^0),\label{sigmap0}\nonumber\\
  -(c_{\mathrm p_{ijkh}}e_{y_{kh}}(\vec\pi)+\alpha_{\mathrm p_{ij}}) p_\mathrm f^0,\\
  \sigma_{\mathrm c_{ij}}^0=
  (c_{\mathrm c_{ijlm}}e_{y_{lm}}(\vec\omega^{kh})+c_{\mathrm c_{ijkh}})e_{x_{kh}}(\vec u^0),\label{sigmac0}.
\end{numcases}
\subsubsection{Boundary value problem for $\vec v_\mathrm p^0$ and $p_\mathrm f^1 $}
We now consider Eq.(\ref{eqp3}) at ${\mathcal O} (\varepsilon^{-1}) $, Eq.(\ref{eqp4}) at ${\mathcal O} (\varepsilon^{O}) $, and boundary condition Eq.(\ref{eqg3}) at ${\mathcal O}(\varepsilon^0) $, from which we deduce the differential system
\begin{numcases}{}
\dfrac{\partial}{\partial y_i}\left[ \dfrac{K_{\mathrm p_{ij}}}{\mu} (\dfrac{\partial p_\mathrm f^1}{\partial y_j} + \dfrac{\partial p_\mathrm f^0}{\partial x_j}) \right]=0
\ \hbox{in $\Omega_\mathrm p,$}\\
\strut\left[ \dfrac{K_{\mathrm p_{ij}}}{\mu} (\dfrac{\partial p_\mathrm f^1}{\partial y_j} + \dfrac{\partial p_\mathrm f^0}{\partial x_j}) \right]n_i=0\ \hbox{over $\Gamma$},\\
p_\mathrm f^1: \hbox{periodic in $\vec y$.}
\end{numcases}
The above set of equations is a well-posed boundary value problem of unknown $p_\mathrm f^1 $, from which it appears that $p_\mathrm f^1$ is a linear function of $\vec\nabla_x p_f^0 $:
\begin{equation}
 p_\mathrm f^1= \chi_{\mathrm p_i}\ \dfrac{\partial p_\mathrm f^0}{\partial x_i} + \bar p_\mathrm f^1 (\vec x, t),
 \label{p1}
\end{equation}
where $\bar p_\mathrm f^1 (\vec x, t)$ is an arbitrary function and where
\begin{equation}
 <\vec\chi_\mathrm p >_{\Omega}^{\Omega_\mathrm p}=\vec 0.
\end{equation}
Vector $\chi_{\mathrm p_k}$ is the specific solution for $p_\mathrm f^1$, corresponding to $
{\partial p_\mathrm f^0}/{\partial x_j}=\delta_{jk}$.
Then, by Eq.(\ref{eqp4}) at ${\mathcal O} (\varepsilon^{-1}) $
\begin{equation}
 v_{\mathrm p_i}^0=-\dfrac{K_{\mathrm p_{ij}}}{\mu} (\dfrac{\partial p_\mathrm f^1}{\partial y_j} + \dfrac{\partial p_\mathrm f^0}{\partial x_j}),
\end{equation}
and by Eq.(\ref{p1}),  we deduce the following expression for $\vec v_\mathrm p^0$:
 \begin{equation}
v_{\mathrm p_i}^0= -\dfrac{K_{p_{ij}}}{\mu}(\dfrac{\partial \chi_k}{\partial y_j}+\delta_{jk})\dfrac{\partial p_\mathrm f^0}{\partial x_k},
\label{vp0}
\end{equation}
where $\tilde \delta$ denotes Kronecker's symbol.
\subsubsection{Macroscopic momentum balance}
Let now consider Eqs.(\ref{eqp1}), (\ref{eqc1}), (\ref{eqg1}) at the third order:
\begin{numcases}{}
 \dfrac{\partial\sigma_{\mathrm p_{ij}}^1}{\partial y_j} + 
 \dfrac{\partial\sigma_{\mathrm p_{ij}}^0}{\partial x_j}=0
 &within $\Omega_\mathrm p,$\label{eqp1_3}\\
 \dfrac{\partial\sigma_{\mathrm c_{ij}}^1}{\partial y_j} + 
 \dfrac{\partial\sigma_{\mathrm c_{ij}}^0}{\partial x_j}=0
 &within $\Omega_\mathrm c,$\label{eqc1_3}\\
 \sigma_{\mathrm p_{ij}}^1\ n_j=\sigma_{\mathrm c_{ij}}^1\ n_j
 &over $\Gamma$\label{eqg1_3}.
\end{numcases}
The homogenisation procedure consists now in integrating Eqs.(\ref{eqp1_3}) and (\ref{eqc1_3}) over $\Omega_\mathrm p$ and $\Omega_\mathrm c$, respectively.
This leads to a compatibility condition, {\em i.e.} a necessary and
sufficient condition for the existence of
solutions for $\vec u_\mathrm s^2$ and $\vec u_\mathrm c^2$, which further represents the first order macroscopic momentum balance. Invoking Gauss' theorem, integration of Eq.(\ref{eqp1_3}) over $\Omega_\mathrm p$ yields
\begin{equation}
 \dfrac{1}{\mid\Omega\mid}\int_{\Gamma}\ \sigma_{\mathrm p_{ij}}^1\ n_j\  dS
 +\dfrac{1}{\mid\Omega\mid}\int_{\Omega_\mathrm p}\ \dfrac{\partial\sigma_{\mathrm p_{ij}}^0}{\partial x_j}\ d\Omega = 0,
 \label{eq1_3_int}
\end{equation}
where the contribution over the cell boundaries, $\delta\Omega\cap\delta\Omega_\mathrm p$, cancel due to $\vec y$-periodicity. 
Then, employing Eq.(\ref{eqg1_3}) and Gauss' theorem, we get
\begin{equation}
 \dfrac{1}{\mid\Omega\mid}\int_{\Gamma}\ \sigma_{\mathrm p_{ij}}^1 n_j d\Omega=
 \dfrac{1}{\mid\Omega\mid}\int_{\Omega_\mathrm c}\dfrac{\partial\sigma_{\mathrm c_{ij}}^0}{\partial x_j}d\Omega.
\end{equation}
Finally, Eq.(\ref{eq1_3_int}) becomes
\begin{equation}
 \dfrac{\partial <\sigma_{\mathrm p_{ij}}^0 >^{\Omega_\mathrm p}}{\partial x_i}+\dfrac{\partial <\sigma_{\mathrm c_{ij}}^0 >^{\Omega_\mathrm c}}{\partial x_i}=0.
 \label{MomBal_0}
\end{equation}
Let us define the total stress $\tilde\sigma_{\mathrm T}$ as
\begin{numcases}{\tilde\sigma_{\mathrm T}=}
 \tilde\sigma_\mathrm p&in $\Omega_\mathrm p$,\nonumber\\
 \tilde\sigma_\mathrm c&in $\Omega_\mathrm c$.\nonumber
\end{numcases}
Thus, Eq.(\ref{MomBal_0}) is rewritten as
\begin{numcases}{}
 \dfrac{\partial <\sigma_{\mathrm T_{ij}}^0 >^{\Omega}}{\partial x_i}=0,
\label{Macro_Eq1}\\
<\sigma_{\mathrm T_{ij}}^0 >^{\Omega}=
 <\sigma_{\mathrm p_{ij}}^0 >^{\Omega_\mathrm p}
 +<\sigma_{\mathrm c_{ij}}^0 >^{\Omega_\mathrm c}.
\end{numcases}
Finally, using Eqs.(\ref{sigmap0})-(\ref{sigmac0}), we  get
\begin{equation}
 <\sigma_{\mathrm T_{ij}}^0 >^{\Omega}=C_{ijkh}^{\mathrm{eff}}\ e_{x_{kh}}(\vec u^0)-A_{ij}^{\mathrm{eff}}\ p_\mathrm f^0,
 \label{Macro_Eq2}
\end{equation}
with
\begin{numcases}{}
C^{\mathrm{eff}}_{ijkh}= <c_{\mathrm p_{ijlm}}e_{y_{lm}}(\vec\omega_{\mathrm p}^{kh})+c_{\mathrm p_{ijkh}}>_{\Omega}^{\Omega_\mathrm p}\nonumber\\
+ <c_{\mathrm c_{ijlm}}e_{y_{lm}}(\vec\omega_{\mathrm c}^{kh})+c_{\mathrm c_{ijkh}}>_{\Omega}^{\Omega_\mathrm c},\label{C}\\
A^{\mathrm{eff}}_{ij}=<c_{p_{ijkh}}e_{y_{kh}}(\vec\pi_{\mathrm p})+\alpha_{p_{ij}}>_{\Omega}^{\Omega_\mathrm p}\label{A}.
\end{numcases}
The first order momentum balance is thus described by Eqs.(\ref{Macro_Eq1}), (\ref{Macro_Eq2}).
\subsubsection{Macroscopic mass balance}
At the second order, Eqs.(\ref{eqp3}) and (\ref{eqg3}) yield
\begin{numcases}{}
 \dfrac{\partial v_{\mathrm p_i}^1}{\partial y_i}
 +\dfrac{\partial v_{\mathrm p_i}^0}{\partial x_i}=
 - \alpha_{\mathrm p_{ij}}\left[e_{y_{ij}} (\dfrac{\partial \vec u_\mathrm s^1}{\partial t})
 +e_{x_{ij}}(\dfrac{\partial \vec u^0}{\partial t})\right] \nonumber\\ -
  \beta_\mathrm p \dfrac{\partial p_\mathrm f^0}{\partial t}
 \ \hbox{within $\Omega_\mathrm p $,}\label{eqp3_2}\\
 v_{\mathrm p_i}^1\ n_i=0
 \hspace{0,5cm}\hbox{over $\Gamma$.}\label{eqg3_2}
\end{numcases}
Integrating Eq.(\ref{eqp3_2}) and invoking Gauss' theorem, while taking boundary condition Eq.(\ref{eqg3_2}) into account, together with the condition of periodicity, yields
\begin{equation}
\begin{array}{l}
  \dfrac{\partial < v_{\mathrm p_i}^0>^{\Omega_\mathrm p}}{\partial x_i}=\vspace{0,2cm}\\
  - <\alpha_{\mathrm p_{ij}}(e_{y_{ij}} (\dfrac{\partial \vec u_\mathrm s^1}{\partial t})+e_{x_{ij}}(\dfrac{\partial \vec u^0}{\partial t}) >_{\Omega}^{\Omega_\mathrm p} \vspace{0,2cm}\\-
  <\beta_\mathrm p >_{\Omega}^{\Omega_\mathrm p}\dfrac{\partial p_\mathrm f^0}{\partial t}.
  \end{array}
  \label{macro_mass_bal}
\end{equation}
Using Eq.(\ref{us1}), the above equation can be written as
\begin{equation}
  \dfrac{\partial < v_{\mathrm p_i}^0>^{\Omega_\mathrm p}}{\partial x_i}=- G_{lm}^{\mathrm{eff}}\ e_{x_{lm}}(\dfrac{\partial \vec u^0}{\partial t}) -B^{\mathrm{eff}}\ \dfrac{\partial p_\mathrm f^0}{\partial t},
  \label{Macro_Eq3}
\end{equation}
where
\begin{numcases}{}
 G_{lm}^{\mathrm{eff}}=<\alpha_{p_{ij}} e_{y_{ij}}(\vec\omega_\mathrm p^{lm}) + \alpha_{p_{lm}}>^{\Omega_\mathrm p},\label{G}\\
 B^{\mathrm{eff}}=<\beta_p -\alpha_{p_{ij}}e_{y_{ij}}(\vec\pi_\mathrm p)>^{\Omega_\mathrm p}\label{B}.
\end{numcases}
Now, by Eq.(\ref{vp0}), we get
\begin{numcases}{}
  <v_{\mathrm p_i}^0>^{\Omega_\mathrm p}= - \dfrac{K_{ij}^{\mathrm{eff}}}{\mu}\dfrac{\partial p_\mathrm f^0}{\partial x_j},
   \label{Macro_Eq4}\\
    K_{ik}^{\mathrm{eff}}=<K_{\mathrm p_{ij}}(\dfrac{\partial \chi_k}{\partial y_j}+\delta_{jk}) >^{\Omega_\mathrm p}.
    \label{K}
\end{numcases}
The first order macroscopic mass balance is thus given by Eqs.(\ref{Macro_Eq3}) and (\ref{Macro_Eq4}).
\section{Macroscopic description}
\label{macro_desc}
The first order macroscopic description thus consists of by Eqs.(\ref{Macro_Eq1}), (\ref{Macro_Eq2}), (\ref{Macro_Eq3}) and (\ref{Macro_Eq4}), with the effective properties defined by Eqs.(\ref{C}), (\ref{A}), (\ref{G}), (\ref{B}) and (\ref{K}). From its definition, it is clear that 
tensor $C_{ijkh}^{\mathrm{eff}}$ is the effective elastic tensor of the whole empty medium, made of the empty porous matrix and the inclusions. It therefore coincides with the effective elasticity tensor that would be obtained for a two-phase elastic composite and thus possesses all the required symmetry properties that characterise an elastic tensor ({\em{e.g.}} \cite{Pen17}).  Now, from the varia\-tional formulation associated with the definitions of $\vec u_\mathrm s^1$ and  $\vec u_\mathrm c^1$, we show that the coupling tensors $\tilde A^{\mathrm{eff}}$ and $\tilde G^{\mathrm{eff}}$  are equal ({\em Cf.} \ref{eq_AG}), and that the bulk modulus $B^{\mathrm{eff}}$ is positive ({\em Cf.} \ref{B_pos}). Furthermore,  from the symmetries of $\tilde c_\mathrm p$ and of $\tilde\alpha_\mathrm p$, it follows that $\tilde A^{\mathrm{eff}}$ is symmetric.
With the above mentioned properties, the macroscopic des\-cription, Eqs.(\ref{Macro_Eq1}), (\ref{Macro_Eq2}), (\ref{Macro_Eq3}) and (\ref{Macro_Eq4}), is a  Biot diphasic model of poroelasticity, but in which the effective properties are strongly impacted by the local distribution of inclusions.
An illustration of this appears when considering homogeneous materials. The definitions of the effective Biot parameters and permeability then reduce to
\begin{numcases}{}
 A^{\mathrm{eff}}_{ij}=c_{p_{ijkh}}<e_{y_{kh}}(\vec\pi_{\mathrm p})>^{\Omega_\mathrm p}+ (1-n_\mathrm c)\alpha_{p_{ij}},\nonumber\\
 B^{\mathrm{eff}}=(1-n_\mathrm c)\beta_\mathrm p-\alpha_{p_{ij}}<e_{y_{ij}}(\vec\pi_\mathrm p)>^{\Omega_\mathrm p},\nonumber\\
 K_{ij}^{\mathrm{eff}}=K_{\mathrm p_{ik}}< \dfrac{\partial \chi_j}{\partial y_k}+\delta_{kj} >^{\Omega_\mathrm p},\nonumber
\end{numcases}
where $n_\mathrm c={\mid\Omega_\mathrm c\mid}/{\mid\Omega\mid} $ denotes the inclusion vo\-lume fraction.
We firstly note that at low inclusion concentration, {\em{i.e.}} when $n_\mathrm c \approx 0 $, the inclusions still have an impact since  $\tilde A^{\mathrm{eff}}\neq \tilde\alpha_\mathrm p$ and  $B^{\mathrm{eff}}\neq \beta_\mathrm p$. 
Furthermore, when the porous matrix is incompressible, {\em{i.e}} when $\tilde \alpha_\mathrm p^{\mathrm{eff}}=\tilde I$ and $\beta_\mathrm p^{\mathrm{eff}}=0$, the whole poroelastic composite remains compressible since $\tilde A^{\mathrm{eff}}\neq \tilde I$ and  $B^{\mathrm{eff}}\neq 0$, even at low inclusion vo\-lume fraction. 
Finally, we see that the permeability is such that $K_{ij}^{\mathrm{eff}}=K_{\mathrm p_{ik}}T_{\mathrm p_{kj}}$, where
$$
T_{\mathrm p_{kj}}=<\dfrac{\partial \chi_j}{\partial y_k}+\delta_{kj}>^{\Omega_\mathrm p}.
$$
Tensor $\tilde T_\mathrm p $ is a purely geometrical parameter, and we note that $\phi^{-1} \tilde T_\mathrm p $ actually represents the tortuosity ({\em e.g.} see \cite{Roy10}) associated with the distribution of inclusions.
\section{Conclusion}
\label{conclu}
We have thus shown that the first order macroscopic behaviour, {\em i.e.} with precision in the order of ${\mathcal O}(\varepsilon)$, of a poroelastic matrix containing elastic inclusions is  described by  Biot's diphasic model of poroelasticity
\begin{numcases}{}
 \dfrac{\partial <\sigma_{\mathrm T_{ij}} >^{\Omega}}{\partial x_i}=0,\nonumber\\
  <\sigma_{\mathrm T_{ij}} >^{\Omega}=C_{ijkh}^{\mathrm{eff}}\ e_{x_{kh}}(\vec u)-A_{ij}^{\mathrm{eff}}\ p_\mathrm f, \nonumber\\
   \dfrac{\partial < v_{\mathrm p_i}>^{\Omega_\mathrm p}}{\partial x_i}=- A_{lm}^{\mathrm{eff}}\ e_{x_{lm}}(\dfrac{\partial \vec u}{\partial t}) -B^{\mathrm{eff}}\ \dfrac{\partial p_\mathrm f}{\partial t},\nonumber
   \\
   <v_{\mathrm p_i}>^{\Omega_\mathrm p}=
   \ - \dfrac{K_{ij}^{\mathrm{eff}}}{\mu}\dfrac{\partial p_\mathrm f}{\partial x_j}.\nonumber
\end{numcases}
The effective parameters are strongly affected by the distribution of inclusions, even at low volume fraction. 
In the above developments, it is implicitely assumed that the inclusion size, $l_c$, is of same order of magnitude as the distance between two inclusions. Considering the case of  low inclusion concentration, $l_c \ll l$, would not modify the macroscopic behaviour. But, since this introduces the additional small parameter $l_c / l \ll 1$, simplified formulas can be obtained  for the effective parameters.
% A particular case of the model is derived and solved in \cite{Chen19}, in which an incompressible porous matrix reinforced with isotropic fibers is considered to model a construct for tissue engineering.
While sometimes qualified as a multiporous medium and although three distinct scales are actually involved, this composite is distinct from a double porosity microstructure ({\em{e.g.}} \cite{Aur93}, \cite{Roy12}, \cite{Bou15}), as the distribution of the inclusions does not play the role of a porous network. Indeed, the impact of inclusions on  fluid flow transfer is characterised by a purely geometrical parameter, with no reference to their fluid conductivity.
Finally, we shall underline that all the  above results are valid for elastic properties of both constituents in the same order of magnitude and for perfect interface bonding. A particular case of the model is derived and solved in \cite{Chen19}, in which an incompressible istropic porous matrix reinforced with isotropic fibers is considered to model a construct for tissue engineering. Further work should include numerical simulations on specific geometries, so as to ana\-lyse the sensitivity of inclusion concentration.
\appendix
\section{Boundary value problem for $\vec u_\mathrm s^1$ and $\vec u_\mathrm c^1$}
\label{bvpu1}
Let us multiply the system Eqs.(\ref{pb1_eq1})-(\ref{pb1_eq5}) by a vectorial test function $\vec\gamma$, and then let us integrate over $\Omega$. We obtain the following variational formulation:
\begin{equation}
 \begin{array}{l}
   \displaystyle\int_{\Omega_\mathrm p}c_{\mathrm p_{ijlm}}e_{y_{lm}}(\vec u_\mathrm s^1)e_{y_{ij}}(\vec \gamma)\ d\Omega\\
   + \displaystyle\int_{\Omega_\mathrm c}c_{\mathrm c_{ijlm}}e_{y_{lm}}(\vec u_\mathrm c^1)e_{y_{ij}}(\vec \gamma)\ d\Omega=\vspace{0,2cm}\\
   -  \displaystyle\int_{\Omega_\mathrm p}c_{\mathrm p_{ijlm}}e_{y_{ij}}(\vec \gamma)\ d\Omega\ e_{x_{lm}}(\vec u^0)\vspace{0,2cm}\\
   -\displaystyle\int_{\Omega_\mathrm c}c_{\mathrm c_{ijlm}}e_{y_{ij}}(\vec \gamma)\ d\Omega\ e_{x_{lm}}(\vec u^0)\vspace{0,2cm}\\
   +\displaystyle\int_{\Omega_\mathrm p}\alpha_{p_{ij}}e_{y_{ij}}(\vec\gamma)\ d\Omega\ p_\mathrm f^0,
 \end{array}
 \label{form_var_u1}
\end{equation}
from which it appears that $\vec u_\mathrm s^1$ is a linear vectorial function of $\tilde e_x (\vec u^0) $ and  $ p_f^0$, Eq.(\ref{us1}), and that $\vec u_\mathrm c^1$ is as a linear vectorial function of $\tilde e_x (\vec u^0) $, Eq.(\ref{uc1}).
Third-order tensors $\omega_{\mathrm p_i}^{kh}$ and $\omega_{\mathrm c_i}^{kh}$ are the specific solutions, $\vec u_\mathrm s^1=\vec \omega_\mathrm p^{kh}$, $\vec u_\mathrm c^1=\vec \omega_\mathrm c^{kh}$, to system Eqs.(\ref{pb1_eq1})-(\ref{pb1_eq5}),
corresponding to
\begin{numcases}{}
 e_{x_{lm}}(\vec u^0)=
 \dfrac{1}{2}\left(\delta_{lk}\delta_{mh}+\delta_{mk}\delta_{lh}\right),\nonumber\\
 p_\mathrm f^0=0.\nonumber
\end{numcases}
As for vector $\vec \pi_\mathrm p $, it is the specific solution for $\vec u_\mathrm s^1$ when $p_\mathrm f^0=-1$ and $e_{x_{lm}}(\vec u_s^0)=0$.
\section{Equality of coupling tensors \\$\tilde A^{\mathrm{eff}}$ and $\tilde G^{\mathrm{eff}}$}
\label{eq_AG}
By taking $\vec u_\mathrm s^1=\vec \omega_\mathrm p^{kh}$,  $\vec u_\mathrm c^1=\vec \omega_\mathrm c^{kh}$, $\vec \gamma=\vec \pi_\mathrm p$ in 
$\Omega_\mathrm p$, and  $\vec \gamma=\vec 0$ in $\Omega_\mathrm c$,
in the variational formulation Eq.(\ref{form_var_u1}), we get
\begin{equation}
\begin{array}{l}
   \displaystyle\int_{\Omega_\mathrm p}c_{\mathrm p_{ijlm}}e_{y_{lm}}(\vec \omega_\mathrm p^{kh})e_{y_{ij}}(\vec \pi_\mathrm p)\ d\Omega=\vspace{0,2cm}\\
   -  \displaystyle\int_{\Omega_\mathrm p}c_{\mathrm p_{ijkh}}e_{y_{ij}}(\vec \pi_\mathrm p)\ d\Omega.
   \end{array}
   \label{equalAG_eq1}
\end{equation}
Next, we take $\vec u_\mathrm s^1=\vec\pi_\mathrm p $ and $\vec\gamma=\vec\omega_\mathrm p^{kh}$ in Eq.(\ref{form_var_u1}): 
\begin{equation}
\begin{array}{l}
  \displaystyle\int_{\Omega_\mathrm p}c_{\mathrm p_{ijlm}}e_{y_{lm}}(\vec \pi_\mathrm p)e_{y_{ij}}(\vec\omega_\mathrm p^{kh})\ d\Omega=\vspace{0,2cm}\\
  -\displaystyle\int_{\Omega_\mathrm p}\alpha_{p_{ij}}e_{y_{ij}}(\vec\omega_\mathrm p^{kh})\ d\Omega.
 \end{array}
 \label{equalAG_eq2}
\end{equation}
The left hand sides of Eqs.(\ref{equalAG_eq1}) and (\ref{equalAG_eq2}) are equal. Then, from the equality of both right hand sides, it appears by Eqs.(\ref{A}), (\ref{G}), that $\tilde A^{\mathrm{eff}} =\tilde G^{\mathrm{eff}} $.
\section{Positiveness of the effective Biot bulk modulus $B^{\mathrm{eff}}$}
\label{B_pos}
Considering $\vec u_\mathrm s^1 = \vec \pi_\mathrm p$, $\vec\gamma= \vec \pi_\mathrm p$, in Eq.(\ref{form_var_u1}), we get:
$$
\begin{array}{l}
  \displaystyle\int_{\Omega_\mathrm p}c_{\mathrm p_{ijlm}}e_{y_{lm}}(\vec \pi_\mathrm p)e_{y_{ij}}(\vec\pi_\mathrm p) d\Omega=\\
  -\displaystyle\int_{\Omega_\mathrm p}\alpha_{p_{ij}}e_{y_{ij}}(\vec\pi_\mathrm p) d\Omega.
  \end{array}
$$
The left hand side of the above equation is positive, due to the posi\-tiveness of the local strain energy. Then, from the positiveness of the right hand side, it follows by Eq.(\ref{B}), that $B^{\mathrm{eff}}>0$.
\section*{Acknowledgements}
This research is supported by CNRS (AAP ``Osez l'Interdisciplinarit\'e 2018'', MoTiV Project). P. Recho and C. Verdier are members of LabeX Tec 21 (PIA : Grant No ANR-11-LABEX-0030).

\end{document}